\documentstyle{article}
\newcommand{\hide}[1]{}
\begin{document}

\markboth{{\bf A stable massive charged particle
}} {{G. Rajasekaran}} 

\begin{center}
{\large{\bf A STABLE MASSIVE CHARGED PARTICLE}}\\
\vskip0.5cm
{\bf G. Rajasekaran}
\vskip0.35cm
{\it Institute of Mathematical Sciences, Chennai 600113, India\\
and\\
Chennai Mathematical Institute, Siruseri 603103, India\\
e-mail: graj@imsc.res.in}
\vskip0.35cm
\end{center}

\vspace{1.5cm}
\underline{Abstract} : We consider the possibilty of the existence
of a stable massive charged particle by a minimal extension of
the standard model particle content. Absolute stability in the 
case of singly charged particle is not possible if the usual doublet 
Higgs exists, unless a dicrete symmetry is imposed. But a doubly 
charged particle is absolutely stable.\\

\newpage

Standard Model (SM) of high energy physics based on the gauge group
SU(3)xSU(2)xU(1) has quarks which transform nontrivially under
all the three factor groups SU(3), SU(2) and U(1) while leptons transform
nontrivially under two of them: SU(2) and U(1). Is it possible to
have a fermion that transforms nontrivially under U(1) only? The
answer is yes. It will be an electrically charged fermion that
does not have weak decays or strong interactions. If in addition it is
stable, it will be an interesting particle. It is remarkable that
SM allows such a possibility and the aim of this note is to point
this out.\\

In the SM, the electrical charge Q is given by

$$ Q = I_{3} + \frac {Y}{2} $$

where $ I_{3} $ and Y are the usual SU(2) and U(1) quantum numbers respectively.
We now envisage a fermion $\psi$ which is a singlet under SU(3) and SU(2).
Since Y is arbitrary, we consider two possibilities: (1) Y = -2, Q = -1
and Y = -4, Q = -2.\\

All we have to do is to add the following piece $L_{\psi}$ to the
Lagrangian density of SM:

$$ L_{\psi} = i\bar{\psi}\gamma_{\mu}(\partial_\mu - i\frac{g_1}{2} Y B_{\mu})
\psi - m \bar{\psi}\psi $$

where $B_{\mu}$ is the U(1) gauge field, $g_1$ is its coupling constant and
Y is the hypercharge of $\psi$. We take both the helicity components $\psi_{L}$
and $\psi_{R}$ to have the same Y. Note that the mass term $m\bar{\psi}\psi$
is allowed and hence $m$ is arbitrary. After the breaking of SU(2)xU(1), $B_{\mu}$
can be reexpressed in terms of $A_{\mu}$ and $Z_{\mu}$:

$$ B_{\mu} = cos \theta A_{\mu} + sin \theta Z_{\mu} $$

$$ tan \theta = \frac{g_1}{g_2} $$   

$$ g_2 sin \theta = e $$

where $\theta$ is the usual electroweak mixing angle. Hence $L_{\psi}$ becomes
 
$$ L_{\psi} = i\bar{\psi}\gamma_{\mu}(\partial_{\mu}-ieQA_{\mu}-iQg_2 tan\theta sin\theta
Z_{\mu})\psi - m\bar{\psi}\psi. $$

Thus we see that the only couplings involving $\psi $ are $\bar{\psi}\gamma_{\mu}
\psi A_{\mu} $ and $\bar{\psi}\gamma_{\mu}\psi Z_{\mu} $ and so $\psi\bar\psi$
pair can be produced in high energy collisions of $ e^+e^-, q\bar q, qq $
etc, via virtual $ \gamma $ or Z. Standard formulae for the production of
massive charged fermions in leptonic and hadronic colliders exist in the
literature. To prevent the decay of the Z boson into $ \bar{\psi}\psi $, we impose the
bound:

       $$ m > \frac{m_Z}{2}. $$ \\

What about the Higgs sector? We consider case (1) first. 
The only invariant Higgs coupling is of the
type $ f\bar{l}_L\phi^c\psi, $ where $ l_L $ is any one of the three
left-handed leptonic doublets consisting of a neutrino and a charged lepton
and $ \phi^c $ is the charge-conjugate of the Higgs doublet $ (\phi^0,\phi^-) $.
The nonvanishing vacuum expectation value $ <\phi^0> = v $ leads to
non-diagonal mass terms mixing $\psi$ with $ e, \mu $ and $\tau $. Hence in the charged
lepton sector we have a 4x4 mass matrix which has to be diagonalized to get
the physical fermions $ e, \mu, \tau $ and $\psi$. Actually the Lagrangian
given above must include off-diagonal mass terms connecting $\psi$ with 
all the right-handed charged leptons (which are present
even without Higgs and its vacuum
expectation value). These also must be included in the diagonalization
process.\\

In the 3x3 submatrix of the $ e, \mu, \tau $ sector, the physical Higgs boson
H can be added to $v$, but that is not true of the full 4x4 mass matrix. 
Hence although the diagonalization of the mass matrix
eliminates the off-diagonal flavour-changing couplings such as $\mu e H$,
the couplings $\psi e H, \psi \mu H $ and $\psi \tau H$ remain. The real decays
of $\psi$ into H and a charged lepton can be prevented by  imposing the
bound $ m < m_H + m_{\tau} $. However the decay of $\psi$ into a photon
and a charged lepton through one loop diagrams arising from the off-diagonal 
couplings such as $\psi e H $ cannot be forbidden, thus making $\psi$
unstable.\\

We invoke a discrete symmetry $Z_2$ and assign $Z_2 = -1$ for $\psi$ and
+1 for all other particles of SM including Higgs. This will eliminate
the above Higgs coupling to $\psi$ as well as the off-diagonal mass 
terms originally present before symmetry breaking 
and make $\psi$ absolutely stable.
This completes the model based on case (1).\\

In case (2), there is no need of the discrete symmetry $Z_2$. Conservation of the
U(1) hypercharge itself forbids the Higgs coupling of $\psi$ as well as the
off-diagonal mass terms connecting $\psi$ with the right-handed charged
leptons. Thus the doubly charged $\psi$ is automatically stable.\\

If $\psi$ exists, it will have an important application. Muon catalyzed fusion
is a well-studied phenomenon. Negative muons captured in orbits around
d-d and d-t nuclei (d = deuteron, t = triton) lead to close approach of the 
nuclei inducing fusion. But the instability of the muon has sofar prevented
the practical utilization of this phenomenon. Replacing $\mu$ by $\psi$
we gain in two ways: $\psi$ does not decay and further its high mass will
lead to tighter orbit and increased probability of nuclear fusion. Is this
the explanation for the reported "cold fusion", which is as yet not
an established phenomenon?\\

It is possible that $\psi$ particles produced primordially in the very
early Universe are lurking around, waiting to be discovered.
In any case, search for $\psi$ and $\bar{\psi}$ produced in LHC and the
future linear collider may be worthwhile.\\

Neutral $\psi$-onic atoms such as proton + $\psi$ or $ He^4$ + $\psi$
respectively in the case of singly or doubly charged $\psi$ are
candidates for dark matter. Such models have been studied extensively
by Khlopov and collaborators (1). Variations on this theme have also
been proposed. See for instance, Glashow (2).\\

I thank Shrihari Gopalakrishna and Rahul Sinha for discussions,
Sandip Pakvasa and Xerxes Tata for remarks that led to the revised
version of this note and Maxim Yu Khlopov for informing me of the
work done by his group.\\

References:

(1) M. Yu. Khlopov, arXiv:1012.5756[astro-ph-CO]

(2) S. L. Glashow, arXiv:hep-ph/0504287 
\end{document}